\documentclass[preprint,superscriptaddress]{revtex4-1}
\usepackage{amsmath}
\usepackage{amssymb}
\usepackage{cancel}
\usepackage{amsfonts}
\usepackage[T1]{fontenc}
\usepackage{bm}
\usepackage{color}
\usepackage{graphicx}
 \DeclareSymbolFont{operators}{OT1}{cmr}{m}{n}
 \DeclareSymbolFont{letters}{OML}{cmm}{m}{it}
 \DeclareSymbolFont{symbols}{OMS}{cmsy}{m}{n}
 \DeclareSymbolFont{largesymbols}{OMX}{cmex}{m}{n}

\usepackage{hyperref}
\hypersetup{
    bookmarks=true,         
    unicode=false,          
    pdftoolbar=true,        
    pdfmenubar=true,        
    pdffitwindow=false,     
    pdfstartview={FitH},    
    pdftitle={Fluid moments of the Landau collision operator},    
    pdfauthor={Eero Hirvijoki},     
    pdfsubject={Plasma fluid theory},   
    pdfcreator={Eero Hirvijoki},   
    pdfproducer={GNU Make}, 
    pdfkeywords={keyword1} {key2} {key3}, 
    pdfnewwindow=true,      
    colorlinks=true,        
    linkcolor=blue,      
    citecolor=blue,         
    filecolor=blue,      
    urlcolor=blue           
}
\usepackage{array}

\newcommand{\mhe}[1]{{\bf He}_{(#1)}}
\newcommand{\mh}[1]{{\bf H}_{(#1)}}

\makeatother

\begin{document}

\title{Fluid moments of the Landau collision operator}

\author{E. Hirvijoki}
\affiliation{Princeton Plasma Physics Laboratory, Princeton, New Jersey 08543, USA}
\author{M. Lingam}
\affiliation{Department of Astrophysical Sciences, Princeton University, Princeton, New Jersey 08544, USA}
\affiliation{Princeton Plasma Physics Laboratory, Princeton, New Jersey 08543, USA}
\author{D. Pfefferl\'e}
\affiliation{Princeton Plasma Physics Laboratory, Princeton, New Jersey 08543, USA}
\author{L. Comisso}
\affiliation{Department of Astrophysical Sciences, Princeton University, Princeton, New Jersey 08544, USA}
\affiliation{Princeton Plasma Physics Laboratory, Princeton, New Jersey 08543, USA}
\author{J. Candy}
\affiliation{General Atomics, San Diego, California 92186, USA}
\author{A. Bhattacharjee}
\affiliation{Department of Astrophysical Sciences, Princeton University, Princeton, New Jersey 08544, USA}
\affiliation{Princeton Plasma Physics Laboratory, Princeton, New Jersey 08543, USA}

\date{\today}

\begin{abstract}
One important problem in plasma physics is the lack of an
accurate and complete description of Coulomb collisions in associated
fluid models. To shed light on the problem, this Letter introduces an integral identity
involving the multi-dimensional Hermite tensor polynomials and 
presents a method for computing exact expressions for the fluid moments
of the nonlinear Landau collision operator. The proposed
methodology provides a systematic and rigorous means of extending the
validity of fluid models that have an underlying inverse-square force
particle dynamics to weakly collisional and strong flow regimes.
\end{abstract}

\pacs{put Pacs here}%
\keywords{Inverse-square force; Landau collision operator; Rosenbluth
  potentials; weakly collisional plasmas; galactic dynamics; globular
  clusters; planetary rings}
 
\maketitle


Fluid theories are omnipresent in science. They are {\it de facto} representations of macroscopic behaviour and result from a parent kinetic model in which the particle dynamics is governed by the kinetic equation 
\begin{equation}
\frac{d f_{s}}{dt}=C^{ss'}[f_{s},f_{s'}]. 
\end{equation}
Both the free-streaming Vlasov operator $d/dt=\partial/\partial t + \dot{\bm{x}}\cdot\nabla_{\bm{x}}+\dot{\bm{v}}\cdot\nabla_{\bm{v}}$ and the collision operator $C^{ss'}[f_{s},f_{s'}]$ between particle species $s$ and $s'$ should be treated equally but, given the complexity of most collision operators in the kinetic theory~\cite{villani2002}, self-consistently incorporating the collisional effects into fluid models is generally hard to achieve. 

In rarefied gas theory, two very different approaches exist for embedding collisional effects. The Chapman-Enskog procedure~\cite{Chapman:1970} builds upon an expansion of the distribution function with respect to the relative changes in density, temperature, and flow velocity over an interval equal to the average length of free flow. Higher order moments of the distribution function, such as the heat flux, are sought as spatial derivatives of density, flow velocity, and temperature, by means of solving the transport equation recursively. Grad's procedure~\cite{Grad:1949_rarefied_gas_theory}, on the other hand, looks for the solution of the transport equation from a finite dimensional subspace, a method that has proven successful in many fields of physics, mathematics, and engineering. In this case, the distribution function is expanded using a set of orthogonal polynomials, and the expansion coefficients, which are uniquely written in terms of the traditional fluid moments, are treated as individual variables for which a set of differential equations are derived. In plasma physics, Grad's approach is rarely applied and the standard fluid model by Braginskii~\cite{Braginskii:1965} follows the Chapman-Enskog theory. Although exact polynomial expansions of the nonlinear Landau collision operator have been carried out~\cite{JH09} using the so-called total velocity expansion, fluid models typically work better when the Maxwellian envelope of the polynomial expansion has a nonzero flow velocity.

In this Letter, we delineate a procedure for computing exact fluid moments of the nonlinear Landau collision operator in the so-called random velocity expansion (Grad's approach), allowing for an efficient treatment of weak collisions and strong flows. We proceed by first recalling the multivariate Hermite expansion after which we compute the moments of the Landau collision operator. Finally, we provide the algorithm for generating the exact moments recursively in terms of three 1D-integrals. One application of the presented formalism would be to compute transport coefficients with exact expressions for the collisional moments, thereby extending previous work on the subject; see, e.g., Ref.~\cite{helander2005collisional} for an overview of topics related to transport.

Following Grad, we expand the distribution function of species $s$ in terms of \emph{probabilists'} multivariate Hermite tensor polynomials, $\mhe{i}(\bm{y})\equiv(\bm{y}-\nabla_{\bm{y}})^i1$, according to 
\begin{equation}
f_s=\frac{n_s\beta_s^3}{(2\pi)^{3/2}}e^{-\beta_s^2(\bm{v}-\bm{V}_s)^2/2}\sum_{i=0}^{\infty}\frac{1}{i!}\bm{\alpha}_{s(i)}\mhe{i}(\beta_s(\bm{v}-\bm{V}_s)),
\end{equation}
where the coefficients $n_s(\bm{x},t)$, $\beta_s(\bm{x},t)=\sqrt{m_s/T_s(\bm{x},t)}$, and $\bm{V}_s(\bm{x},t)$ refer to the species' density, temperature (mean kinetic energy), and flow velocity. $\bm{\alpha}_{s(i)}\mhe{i}$ is understood as a full contraction of two tensors of rank $i$ into a scalar (sum over repeated Greek indices $(i)$, $(j)$, $(k)$ etc. is assumed, excluding $s$ which is reserved for particle species). Explicitly, the first few polynomials are $\mhe{0}(\bm{y})=1$, $\mhe{1}(\bm{y})=\bm{y}$, and $\mhe{2}(\bm{y})=\bm{y}\bm{y}-\mathbf{I}$. Representation of higher order polynomials requires index notation.

Since $\mhe{i}$ satisfy the orthogonality condition $\int_{\mathbb{R}^D} d\bm{x}\; \mhe{i}(\bm{x}) \mhe{j}(\bm{x})\;{\cal N}(\bm{x},\mathbf{I})=\bm{\delta}_{(i)(j)}$, where ${\cal N}(\bm{x}-\bm{\mu},\bm{\Sigma})$ is the multivariate Normal distribution and $\bm{\delta}_{(i)(j)}$ is the multivariate identity tensor (see Eq.(19) in Ref.~\cite{Grad:1949_polynomials}), the expansion coefficients $\bm{\alpha}_{s(i)}(\bm{x},t)$ are determined by 
\begin{equation}
\bm{\alpha}_{s(i)}=\frac{1}{n_s}\int_{\mathbb{R}^3} d\bm{v}\; f_s\;\mhe{i}(\beta_s(\bm{v}-\bm{V}_s)).
\end{equation}
Explicitly, the first few terms in the series are $\bm{\alpha}_{s(0)}=1$, $\bm{\alpha}_{s(1)}=\bm{0}$, and $\bm{\alpha}_{s(2)}=\mathbf{p}_s/p_s-\mathbf{I}$, with $\mathbf{p}_s$ the pressure tensor and $p_s=\mathrm{Tr}(\bm{p}_s)/3$ the scalar pressure. It is important to notice that for all species $\bm{\alpha}_{s(0)}$ and $\bm{\alpha}_{s(1)}$ are fixed and that $\bm{\alpha}_{s(2)}$, by construction, is traceless. These properties account for using the density, the flow velocity, and the temperature (the scaled trace of the pressure tensor), to parameterize the Maxwellian envelope. Consequently, the coefficients $\bm{\alpha}_{s(i>0)}$ do not carry density,  momentum, nor energy. In the literature, this expansion introduced by Grad is sometimes referred to as the \emph{random velocity expansion}. The alternative, referred to as the \emph{total velocity expansion}, would drop the flow $\bm{V}_s$ from the exponent and the Hermite functions, and include it into $\bm{\alpha}_{s(1)}$. Obviously, the advantage of the random-velocity expansion is that it better captures strong near-Maxwellian flows with less expansion coefficients. 

Grad's moment equations are obtained by computing the moments of the Vlasov and Landau operators according to 
\begin{align}
V_{(k)}^s&\equiv\int_{\mathbb{R}^3} d\bm{v}\; \mhe{k}(\beta_s(\bm{v}-\bm{V}_s))\frac{df_s}{dt},\\
\label{eq:hermiteCollMoment}
C^{ss'}_{(k)}&\equiv\int_{\mathbb{R}^3} d\bm{v}\; \mhe{k}(\beta_s(\bm{v}-\bm{V}_s))C_{ss'}[f_s,f_{s'}],
\end{align}
and then setting $\{V_{(k)}^s=\sum_{s'}C^{ss'}_{(k)}\}_{k=0}^{\infty}$. In other words, the fluid equations are obtained by taking the projection of the kinetic equation onto a specific set of basis functions, and the fluid moments carry the meaning of \emph{expectation values} or \emph{observables}, as in quantum mechanics. Projecting the distribution function on Hermite polynomials is a mathematically consistent operation because the Hermite functions form a complete orthogonal basis for the underlying Sobolev space and, by invoking the Hilbert-Schmidt theorem, it may also be possible to quantify how accurately the fluid description scales with respect to the number of Hermite polynomials included.

Of course, most existing numerical tools that solve the moment equations typically work with \emph{standard} moments. Since there obviously exists a one-to-one mapping between Hermite polynomials and velocity space monomials, it is always possible to express the standard moments in terms of coefficients $\bm{\alpha}_{s(i)}$ and vice-versa. In fact, the fluid equations can readily be expressed on the Vlasov side in terms of traditional moments and on the collision operator side in terms of Hermite moments. For example, the 10-moment equations would read
\begin{align}
\int_{\mathbb{R}^3}d\bm{v}\; \frac{df_s}{dt} & \equiv 0 
\\
\int_{\mathbb{R}^3}d\bm{v}\; m_s\bm{v}\frac{df_s}{dt} & \equiv m_s\beta^{-1}_s\sum_{s'}C^{ss'}_{(1)}
\\
\int_{\mathbb{R}^3}d\bm{v}\; m_s\bm{v} \bm{v}\frac{df_s}{dt} & \equiv m_s\beta^{-2}_s\sum_{s'}C^{ss'}_{(2)}+m_s\beta^{-1}_s\sum_{s'}\left(\bm{V}_{s} C^{ss'}_{(1)}+C^{ss'}_{(1)} \bm{V}_{s}\right)
\end{align}
where the expressions for $C^{ss'}_{(k)}$ are defined in Eq.~(\ref{eq:hermiteCollMoment}), and we have used the fact that $C^{ss'}_{(0)}\equiv 0$. Higher order moment equations are obtained in a similar way, the task being to translate $C^{ss'}_{(k)}$ in terms of the moments $\left\{\int_{\mathbb{R}^3}d\bm{v}\:\bm{v}^{(j)}f_s\right\}_{j=0}^{\infty}$, where $\bm{v}^{(j)}$ is $j$th-order tensor product as in $\bm{v}^{(2)}=\bm{v} \bm{v}$ and $\bm{v}^{(3)}=\bm{v} \bm{v} \bm{v}$, etc. 

In what follows, we express the Hermite moments of the collision operator as $C^{ss'}_{(k)}= C^{ss'}_{(ijk)}  \bm{\alpha}_{s(i)}\bm{\alpha}_{s'(j)}$, where $C^{ss'}_{(ijk)}$ only depends on $\bm{\Delta}_{ss'}\equiv\beta_{s'}(\bm{V}_{s}-\bm{V}_{s'})$ and $\theta_{ss'}\equiv\beta_{s'}/\beta_{s}$. We also derive recursion relations for generating the functions $C^{ss'}_{(ijk)}$ in terms of three 1D-integrals (parametrized by $\bm{\Delta_{ss'}}$ and $\theta_{ss'}=\beta_{s'}/\beta_{s}$). Since $\bm{\alpha}_{s(i)}$ are uniquely defined in terms of the moments $\left\{\int_{\mathbb{R}^3}d\bm{v}\:\bm{v}^{(j)}f_s\right\}_{j=0}^{i}$, all the necessary terms for consistent moment equations are provided. Furthermore, truncating the Hermite series by setting $\{\bm{\alpha}_{s(i)}\}_{i=N+1}^{\infty}=0$ for a given $N$ also uniquely truncates the Vlasov moments and provides a consistent closure for the fluid equations. Consider, for example, the 10-moment equations for which the distribution function is truncated by setting $\bm{\alpha}_{s(i>2)}=0$. This leads to the condition
\begin{equation}
  \int_{\mathbb{R}^3}d\bm{v}\; m_s\;v^iv^jv^k\; f_s=n_sm_sV_s^iV_s^jV_s^k+p_s^{ij}V_s^k+p_s^{jk}V_s^i+p_s^{ki}V_s^j,
\end{equation}
and provides an expression for the third order moment which is required in the dynamical equation for the second moment.

In order to compute the Hermite moments of the Landau collision operator, it is useful to introduce the so-called \emph{physicists'} multi-dimensional Hermite tensor polynomials~\cite{Holmquist:1996}, 
$\mh{i}(\bm{y})=(2\bm{y}-\nabla_{\bm{y}})^i1$, as our derivation heavily relies on the following, apparently unknown, integral identity
\begin{equation}
\label{eq:beautiful_relation}
e^{-\bm{x}^2/2}\mhe{i}(\bm{x})=(2/\pi)^{D/2}\int_{\mathbb{R}^D}d\bm{y}\; e^{-(\bm{x}-\bm{y})^2-\bm{y}^2}\mh{i}(\bm{y}).
\end{equation}
A simple proof is given by induction: the case $i=0$ follows from the convolution of two Gaussian distributions, and the step $(i-1)\rightarrow i$ follows from the recursion relations for $\mhe{i}$ and $\mh{i}$.  By virtue of Eq.~(\ref{eq:beautiful_relation}), the distribution function $f_s$ is then conveniently expressed as
\begin{equation}
f_s=\int_{\mathbb{R}^3}d\bm{u}\;{\cal N}\left(\bm{u}-\bm{v},\mathbf{I}/2\beta^2_s\right)\; g_s,
\end{equation}
where the function $g_s(\bm{x},\bm{u},t)$ is 
\begin{equation}
\label{eq:g}
g_s=n_s\;{\cal N}\left(\bm{u}-\bm{V}_s,\mathbf{I}/2\beta^2_s\right)\sum_{i=0}^{\infty}\frac{1}{i!}\bm{\alpha}_{s(i)} \mh{i}(\beta(\bm{u}-\bm{V}_s)).
\end{equation}
Essentially, $f_s$ is a Gaussian convolution (or Gauss-Weierstrass transform) of $g_s$ and the identity in Eq.~(\ref{eq:beautiful_relation}) establishes  that $f_s$ and $g_s$ are related through the coefficients $\bm{\alpha}_{s(i)}$ in a simple (diagonal) way. Writing $f_s$ as a convolution will help us to evaluate the Landau collision operator because any integral or differential operator acting on $f_s$ is readily transferred to act on the Gaussian kernel. In Ref.~\cite{Hirvijoki20152735} this fact was used to construct a Gaussian radial-basis-function approach for solving the kinetic equation. 

Using the Rosenbluth-MacDonald-Judd-Trubnikov potential formulation~\cite{RMJ1957,Trubnikov1958}, the Landau collision operator can be written as 
\begin{equation}
C_{ss'}[f_s,f_{s'}]\equiv c_{ss'}\nabla_{\bm{v}}\cdot\left[\mu_{ss'}\nabla_{\bm{v}}\phi_{s'} f_s-\nabla_{\bm{v}}\cdot\left(\nabla_{\bm{v}}\nabla_{\bm{v}}\psi_{s'} f_s\right)\right],
\end{equation}
where, in plasmas, $c_{ss'}=\ln\Lambda (e_se_{s'})^2/(m_s\varepsilon_0)^2$, and $\mu_{ss'}=1+m_s/m_{s'}$, with $\ln\Lambda$ indicating the Coulomb Logarithm, $e_s$ the species charge, and $m_s$ the species mass. The potential functions, $\phi_s(\bm{x},\bm{v},t)$, and $\psi_s(\bm{x},\bm{v},t)$, appearing in the collision operator are weighted integrals of the distribution function according to
\begin{equation}
\begin{pmatrix}
\phi_s\\ \psi_s
\end{pmatrix}
=-\frac{1}{4\pi}\int_{\mathbb{R}^3} d\bm{v}' f_s
\begin{pmatrix}
\lvert \bm{v}-\bm{v}'\rvert^{-1}\\
\lvert \bm{v}-\bm{v}'\rvert/2
\end{pmatrix}.
\end{equation}
The integral identity in Eq.~(\ref{eq:beautiful_relation}) transfers the velocity dependency of the distribution function to a Gaussian function in such a way that, when the alternative form for $f_s$ is substituted into the expressions for $\phi_s$ and $\psi_s$, we find
\begin{equation}
\label{eq:transformed_potentials}
\begin{pmatrix}
\phi_s\\
\psi_s
\end{pmatrix}
=-\frac{1}{4\pi}\int_{\mathbb{R}^3} d\bm{u}\; g_s
\begin{pmatrix}
\beta_s\Phi(\beta_s\lvert \bm{v}-\bm{u}\rvert)\\
\Psi(\beta_s\lvert \bm{v}-\bm{u}\rvert)/2\beta_s
\end{pmatrix},
\end{equation}
where $\Phi(z)=\mathrm{erf}(z)/z$ and $\Psi(z)=[z+1/(2z)]\mathrm{erf}(z)+\exp(-z^2)/\sqrt{\pi}$, with $\mathrm{erf}(z)$ the error-function. Although a seemingly difficult $\mathbb{R}^3$-integral remains in the new expressions for $\phi_s$ and $\psi_s$, our transformation already has the computational advantage that the singularity in the integrand for the potential $\phi_s$ has been removed: both $\Phi$ and $\Psi$ are nonsingular.
 
The true benefits of our transformation, however, become apparent when the $\mhe{k}(\beta_s(\bm{v}-\bm{V}_s))$-moments of the collision operator are computed. Using the transformed potentials defined in Eq.~(\ref{eq:transformed_potentials}), substituting the expressions for $f_s$ and $g_s$, multiplying the collision operator with $\mhe{k}(\beta_s(\bm{v}-\bm{V}_s))$, integrating over the velocity space, and further integrating by parts gives an exact expression for $C^{ss'}_{(k)}=\sum_{i,j=0}^{\infty}\bm{\alpha}_{s(i)}\bm{\alpha}_{s'(j)}C^{ss'}_{(ijk)}$ in terms of 
\begin{widetext}
\begin{multline}
\label{eq:result}
C^{ss'}_{(ijk)}=\frac{n_sn_{s'}c_{ss'}}{4\pi i!j!}\int_{\mathbb{R}^3}d\bm{z}\;{\cal N}\left(\bm{z}+\bm{\Delta}_{ss'},\mathbf{I}/2\sigma^2_{ss'}\right)
\Biggr\{\mu_{ss'}\beta_s\beta^2_{s'}\frac{\Phi'(z)}{z}\bm{z}\cdot\mathsf{G}^{\Phi}_{(ijk)}\left(\bm{z}+\bm{\Delta}_{ss'},\theta_{ss'}\right)\\
+\beta_{s'}\beta^2_s\Big[\left(\Psi''(z)-\frac{\Psi'(z)}{z}\right)\frac{\bm{z}\bm{z}}{z^2}+\frac{\Psi'(z)}{z}\mathbf{I}\Big]:\mathsf{G}^{\Psi}_{(ijk)}\left(\bm{z}+\bm{\Delta}_{ss'},\theta_{ss'}\right)\Biggr\}.
\end{multline}
where, for the sake of compactness, we have defined the scalar $\sigma_{ss'}^2=1/(1+2\theta^2_{ss'})$ and the vector $\bm{\Delta}_{ss'} =\beta_{s'}(\bm{V}_{s'}-\bm{V}_s)$. The functions $\mathsf{G}^{\Phi}_{(ijk)}(\bm{c},\theta)$ and $\mathsf{G}^{\Psi}_{(ijk)}(\bm{c},\theta)$, which are rank $i+j+k+1$ and $i+j+k+2$ tensor polynomials of the argument $\bm{c}$, are compactly expressed as the following analytic integrals 
\begin{align}
\label{eq:gphi}
\mathsf{G}^{\Phi}_{(ijk)}(\bm{c},\theta)&\equiv\int_{\mathbb{R}^3}d\bm{y}\;\nabla_{\bm{y}}\mhe{k}(\bm{y}+2\theta\sigma^2\bm{c})\mhe{i}(\bm{y}+2\theta\sigma^2\bm{c})\mh{j}(\theta\bm{y}-\sigma^2\bm{c}){\cal N}(\bm{y},\sigma^2\mathbf{I})
\\
\label{eq:gpsi}
\mathsf{G}^{\Psi}_{(ijk)}(\bm{c},\theta)&\equiv\frac{1}{2}\int_{\mathbb{R}^3}d\bm{y}\; \nabla_{\bm{y}}\nabla_{\bm{y}}\mhe{k}(\bm{y}+2\theta\sigma^2\bm{c}) \mhe{i}(\bm{y}+2\theta\sigma^2\bm{c})\mh{j}(\theta\bm{y}-\sigma^2\bm{c}){\cal N}(\bm{y},\sigma^2\mathbf{I}).
\end{align}
\end{widetext}
Notice that the dot and double dot products in Eq.~(\ref{eq:result}) are to be contracted with the respective nablas appearing in Eqs.~(\ref{eq:gphi}) and~(\ref{eq:gpsi}). 

It is useful to notice that the scalar functions $\Phi'(z)/z$, $\Psi'(z)/z$, and $\Psi''(z)-\Psi'(z)/z$ appearing in Eq.~(\ref{eq:result}) are bounded and that $\lim_{z\rightarrow 0}\Psi''(z)-\Psi'(z)/z = 0$. These conditions guarantee that the integrand of $C^{ss'}_{(ijk)}$ has no singular points. It should also be realized that no matter at what order the series of Hermite polynomials is truncated, the momentum and energy balance of the resulting fluid equations are always consistent: for any given distribution function in this representation of the collision operator, exact conservation properties are guaranteed.

As such, Eq.(\ref{eq:result}) is a complex object involving two integrals over $\mathbb{R}^3$, but, as shown hereafter, it is possible to reduce it to a finite sum of tensors derived from a recursion relation involving three 1D-integrals. The final result is then exact and programmable.

Dropping the $s$ labels for conciseness, we notice that the functions $\mathsf{G}_{(ijk)}^{\Phi}(\bm{z}+\bm{\Delta},\theta)$ and $\mathsf{G}_{(ijk)}^{\Psi}(\bm{z}+\bm{\Delta},\theta)$ as well as the Normal distribution ${\cal N}\left(\bm{z}+\bm{\Delta},\mathbf{I}/2\sigma^2\right)$ in Eq.~(\ref{eq:result}) depend on the same argument $\bm{z}+\bm{\Delta}$. We expand $\bm{z}=(\bm{z}+\bm{\Delta})-\bm{\Delta}$, so that
\begin{align}
\bm{z}\cdot\mathsf{G}_{(ijk)}^{\Phi}(\bm{z}+\bm{\Delta},\theta)
=(\bm{z}+\bm{\Delta})\cdot\mathsf{G}_{(ijk)}^{\Phi}(\bm{z}+\bm{\Delta},\theta)-\bm{\Delta}\cdot\mathsf{G}_{(ijk)}^{\Phi}(\bm{z}+\bm{\Delta},\theta).
\end{align}
Analogously, the function $\bm{z}\bm{z}:\mathsf{G}_{(ijk)}^{\Psi}(\bm{z}+\bm{\Delta},\theta)$ is written as a sum of terms that depend on $\bm{z}+\bm{\Delta}$ and are contracted with polynomials of $\bm{\Delta}$. Each of those terms, being now polynomial of $\bm{z}+\bm{\Delta}$, is then further decomposed into a sum of Hermite polynomials $\mh{m}(\sigma(\bm{z}+\bm{\Delta}))$. The reason for this extra step is that the probabilists' Hermite polynomials satisfy the generating formula
\begin{equation}
\nabla^m_{\bm{\Delta}}{\cal N}\left(\bm{z}+\bm{\Delta},\mathbf{I}/2\sigma^2\right)= (2\sigma^2)^{m}{\cal N}\left(\bm{z}+\bm{\Delta},\mathbf{I}/2\sigma^2\right)\mh{m}\left(\sigma(\bm{z}+\bm{\Delta})\right),
\end{equation}
so that the products involving $\bm{z}$, $\mathsf{G}$'s and the Normal distribution are efficiently reduced to a sum of $\bm{\Delta}$-derivatives of ${\cal N}\left(\bm{z}+\bm{\Delta},\mathbf{I}/2\sigma^2\right)$. The derivatives with respect to $\bm{\Delta}$ can then be taken outside the integral over $\bm{z}$ in Eq.(\ref{eq:result}) and the remaining $\mathbb{R}^3$-integrals simplified into three 1D-integrals
\begin{align}
{\cal I}^{(0)}(\Delta,\sigma)&\equiv\frac{\sigma}{\Delta\sqrt{\pi}}\int_0^{\infty}dz\; \left[e^{-\sigma^2(z-\Delta)^2}-e^{-\sigma^2(z+\Delta)^2}\right]\Phi'(z),\\
{\cal G}^{(0)}(\Delta,\sigma)&\equiv\frac{\sigma}{\Delta\sqrt{\pi}}\int_0^{\infty}dz\; \left[e^{-\sigma^2(z-\Delta)^2}-e^{-\sigma^2(z+\Delta)^2}\right]\left(\frac{\Psi''(z)}{z}-\frac{\Psi'(z)}{z^2}\right),\\
{\cal H}^{(0)}(\Delta,\sigma)&\equiv\frac{\sigma}{\Delta\sqrt{\pi}}\int_0^{\infty}dz\; \left[e^{-\sigma^2(z-\Delta)^2}-e^{-\sigma^2(z+\Delta)^2}\right]\Psi'(z).
\end{align}
Letting the functions $\{{\cal I}^{(m)}, {\cal G}^{(m)}, {\cal H}^{(m)}\}(\bm{\Delta},\sigma)\equiv (2\sigma^2)^{-m}\nabla^m_{\bm{\Delta}}\{{\cal I}^{(0)}, {\cal G}^{(0)}, {\cal H}^{(0)}\}(\Delta,\sigma)$, the outlined procedure finally leads to 
\begin{widetext}
\begin{align}
\begin{split}
\label{eq:final_result}
C^{ss'}_{(ijk)}=&\frac{n_sn_{s'}c_{ss'}}{4\pi i!j!}\Biggr\{
\mu_{ss'}\beta_s\beta^2_{s'}\Biggr[\sum_{m=0}^{i+j+k}\Gamma^{\Phi}_{(ijk)(m)} {\cal I}^{(m)}\left(\bm{\Delta}_{ss'},\sigma_{ss'}\right)
\\&
\qquad\qquad\qquad\qquad
-\bm{\Delta}_{ss'}\cdot\sum_{m=0}^{i+j+k+1}\Gamma_{(ijk)(m)}^{\Phi1} {\cal I}^{(m)}\left(\bm{\Delta}_{ss'},\sigma_{ss'}\right)\Biggr] 
\\&
\qquad\qquad+\beta_{s'}\beta^2_s\Biggr[\sum_{m=0}^{i+j+k}\Gamma^{\Psi}_{(ijk)(m)} {\cal G}^{(m)}\left(\bm{\Delta}_{ss'},\sigma_{ss'}\right)
\\&
\qquad\qquad\qquad\qquad
-2\bm{\Delta}_{ss'}\cdot\sum_{m=0}^{i+j+k+1}\Gamma_{(ijk)(m)}^{\Psi1} {\cal G}^{(m)}\left(\bm{\Delta}_{ss'},\sigma_{ss'}\right)
\\&
\qquad\qquad\qquad\qquad+\bm{\Delta}_{ss'}\bm{\Delta}_{ss'}:\sum_{m=0}^{i+j+k+2}\Gamma_{(ijk)(m)}^{\Psi2} {\cal G}^{(m)}\left(\bm{\Delta}_{ss'},\sigma_{ss'}\right)
\\&
\qquad\qquad\qquad\qquad
+\mathbf{I}:\sum_{m=0}^{i+j+k+2}\Gamma^{\Psi2}_{(ijk)(m)} {\cal H}^{(m)}\left(\bm{\Delta}_{ss'},\sigma_{ss'}\right)\Biggr]\Biggr\},
\end{split}
\end{align}
where the coefficients $\Gamma_{(ijk)(m)}$ are determined by expanding the algebra or by projecting against the appropriate Hermite function according to 
\begin{equation}
\begin{pmatrix}
\Gamma^{\Phi}_{(ijk)(m)}
\\
\Gamma^{\Phi1}_{(ijk)(m)}
\\
\Gamma^{\Psi}_{(ijk)(m)}
\\
\Gamma^{\Psi1}_{(ijk)(m)}
\\
\Gamma^{\Psi2}_{(ijk)(m)}
\end{pmatrix}
\equiv\int_{\mathbb{R}^3}d\bm{c}
\begin{pmatrix}
\bm{c}\cdot\mathsf{G}^{\Phi}_{(ijk)}\left(\bm{c},\theta_{ss'}\right)
\\
\mathsf{G}^{\Phi}_{(ijk)}\left(\bm{c},\theta_{ss'}\right)
\\
\bm{c}\bm{c}:\mathsf{G}^{\Psi}_{(ijk)}\left(\bm{c},\theta_{ss'}\right)
\\
\bm{c}\cdot\mathsf{G}^{\Psi}_{(ijk)}\left(\bm{c},\theta_{ss'}\right)
\\
\mathsf{G}^{\Psi}_{(ijk)}\left(\bm{c},\theta_{ss'}\right)
\end{pmatrix}
 \mh{m}(\sigma_{ss'}\bm{c}){\cal N}(\bm{c}; \mathbf{I}/2\sigma_{ss'}^2),
\end{equation}
\end{widetext}

As a final note, we point out that since, e.g., ${\cal I}^{(m)}(\bm{\Delta},\sigma)$ is a $\bm{\Delta}$-derivative of the function ${\cal I}^{(0)}(\Delta,\sigma)$ which depends only on the norm $\Delta$, evaluation of ${\cal I}^{(m)}(\bm{\Delta},\sigma)$ involves only $m$ different 1D-integrals, although $\{{\cal I}^{(m)}, {\cal G}^{(m)}, {\cal H}^{(m)}\}(\bm{\Delta},\sigma)$ are tensors of rank $m$. 

In this Letter, exact fluid moments of the nonlinear Landau collision operator are derived by exploiting an apparently unknown integral identity involving multi-dimensional Hermite tensor polynomials. The calculations are carried out without approximations, and rigorously provide a means of extending the validity of fluid models, with underlying inverse-square force particle dynamics, to weakly collisional and strong flow regimes. We have further shown that the moments of the collision operator can be expressed as a sum of 1D-integrals which are generated by three functions. From the computational point-of-view, this guarantees an algorithmic approach. Furthermore, our procedure can be adapted to obtain fluid moments of the collision operator in other fields of physics, such as Newtonian gravitational dynamics~\cite{BinneyTremaine87}. 

{\it Acknowledgments --} The authors are grateful to Ronald E. Waltz for valuable comments. 


\bibliographystyle{unsrt}
\bibliography{bibfile}

\end{document}